\documentclass[twoside,onecolumn]{article}

\usepackage{blindtext} 

\usepackage[sc]{mathpazo} 
\usepackage[T1]{fontenc} 
\usepackage{microtype} 

\usepackage[english]{babel} 

\usepackage[hmarginratio=1:1,top=22mm,columnsep=20pt]{geometry} 
\usepackage[hang, small,labelfont=bf,up,textfont=it,up]{caption} 
\usepackage{booktabs} 

\usepackage{lettrine} 

\usepackage{enumitem} 
\setlist[itemize]{noitemsep} 

\usepackage{abstract} 

\usepackage{titlesec} 
\renewcommand\thesection{\Roman{section}} 
\renewcommand\thesubsection{\roman{subsection}} 
\titleformat{\section}[block]{\large\scshape\centering}{\thesection.}{1em}{} 
\titleformat{\subsection}[block]{\large}{\thesubsection.}{1em}{} 

\usepackage{fancyhdr} 
\usepackage{titling} 

\usepackage{hyperref} 

\usepackage{natbib}
\usepackage{graphicx}
\usepackage{amsmath}

\pretitle{\begin{center}\Huge\bfseries} 
\posttitle{\end{center}} 
\title{The Fractal Nature of Clouds in Global Storm-Resolving Models} 
\author{%
\textsc{Hannah~M.~Christensen}\thanks{Corresponding author. hannah.christensen@physics.ox.ac.uk} \\[1ex] 
\and 
\textsc{Oliver G. A. Driver} \\[1ex] 
\normalsize Atmospheric, Oceanic and Planetary Physics, University of Oxford, UK. \\ 
}
\date{\today} 
\renewcommand{\maketitlehookd}{%
\begin{abstract}
\noindent 
Clouds in observations are fractals: they show self-similarity across scales ranging from one to 1000 km. This includes individual storms and large-scale cloud structures typical of organised convection. It is not known whether global storm-resolving models reproduce the observed fractal scaling laws for clouds and organised convection. We compute the fractal dimension of clouds using Himawari satellite data and compare this to global storm-resolving model simulations completed as part of the DYAMOND intercomparison project. We find cloud fields in these simulations are indeed fractal, and reproduce the observed fractal dimension to within 10\%. We find the fractal dimension is sensitive to the choice of boundary layer parametrisation scheme used in each model simulation, and not to the convection parametrisation as might have been expected. The fractal dimension is independent of cloud area distributions, providing a complementary metric to assess the multi-scale structure of convection and convective organisation in model simulations.
\end{abstract}
}

\begin{document}

\maketitle


\section*{Plain Language Summary}
Clouds are fractals: they are similar in their appearance whether we consider individual storms, or massive organised cloud systems spanning 1000 km. In this paper, we demonstrate that state-of-the-art high-resolution atmospheric simulations are able to mimic this behaviour. We compute the fractal dimension of clouds, which quantifies how smooth or crinkly clouds are. The fractal dimension of the simulated cloud fields is within 10\% of that observed. We find the fractal dimension is sensitive to choices made when building the computer model used to produce the simulations. Unexpectedly, we find that the choices made when deciding how to represent turbulence close to the surface are important, as opposed to choices made in how to represent thunderstorms. The physics underpinning the merging of clouds into large organised systems is largely unknown. Our paper highlights the ability of simulations to capture this behaviour. This means that high-resolution models can be used as digital laboratories to study the physics underpinning this important process.

\section*{Key Points}
\begin{itemize}
\item Global storm-resolving models are able to reproduce the fractal nature of clouds.
\item The fractal dimension of clouds is used as a model validation tool to assess convective organisation.
\item The fractal dimension is sensitive to boundary layer closure.
\end{itemize}

\section{Introduction}

Numerical models are a useful tool to understand the climate system. Low-resolution climate models are routinely used for long simulations to address questions related to long-term climate change, the physics of low-frequency climate variability, and the likelihood of rare events. However, persistent biases remain in these models \citep[e.g.][]{ar5ch9}. Many biases and inconsistencies between models can be traced to the approximations used when developing model parametrisation schemes \citep[e.g.][]{zelinka2020}.

Recent years have seen an increase in the production of global storm-resolving (also called convection-permitting) atmospheric simulations. Global storm-resolving models directly simulate important processes that must be parametrised in conventional climate models, most notably deep convective clouds \citep{palmer2019}. This substantially improves the fidelity of clouds and precipitation in those models \citep{stevens2020}. This enables us to use these models as a digital laboratory to explore fundamental unsolved questions in atmospheric physics, such as the physics of convective organisation \citep{Wing2019}. 

Before using storm-resolving models in this way, we must quantify and understand the differences and similarities between the models and the real world. This reveals what questions can and cannot be addressed using those models. Certainly, global storm-resolving models appear realistic when compared to observations. An example of this is the visual similarity between images of simulated cloud condensate fields and satellite images of the Earth \citep{stevens2019}. Assessing a model simulation using a visual comparison with observed fields as described here has been coined the `Palmer-Turing test' \citep{Palmer2016} by analogy with the Turing test for Artificial Intelligence. While it can be difficult at first glance to distinguish between model and observed data, on closer inspection, the satellite image can often be identified.

It has long been known that clouds exhibit self-similarity across scales ranging from one to 1000 km \citep{lovejoy1982} --- i.e. they are \emph{fractals}. This behaviour can provide insights into the physics underlying convective aggregation \citep{Haerter2019} and precipitation processes \citep{peters2006}. In particular, scaling relationships imply emergent laws and universal behaviour \citep{lovejoy2013}.  The fractal behaviour of clouds has been used to assess the fidelity of large eddy simulations \citep{siebesma2000}. However, the approach has not been used to assess simulations with domain sizes large enough to simulate convective aggregation. In particular, it is not known whether convection in global storm-resolving models reproduces the observed behaviour. However, in order to use such models to understand the physics of convective aggregation, we must be confident that they represent this process well.

In this paper, we take the Palmer-Turing test as a starting point to assess global storm-resolving models. However, we seek to go further and \emph{quantify} the extent to which a simulated cloud field appears similar to that observed. We achieve this by computing the fractal dimension of clouds. We use this approach to assess the ensemble of global storm-resolving simulations produced for the DYAMOND project. In Section~\ref{sec:data}, we start by describing the model and observational datasets used in this paper. In Section~\ref{sec:methods} we describe how to identify comparable cloud objects in satellite and model data, and the definition of the fractal dimension. In Section~\ref{sec:results} we compare the fractal dimensions thus computed for observed and simulated cloud fields. The DYAMOND models differ from each other in terms of resolution, parametrisation schemes, and other modelling choices: we assess which of these choices impacts convective aggregation as measured by the fractal dimension. Finally we discuss the significance of our results and draw conclusions in Section~\ref{sec:conc}.

\section{Data} \label{sec:data}

\subsection{DYAMOND Global Storm-Resolving Models}

We evaluate the global storm-resolving simulations produced for the DYAMOND summer ensemble \citep{stevens2019}. Each model simulation was initialised at 00:00 UTC on 1st August 2016 and spans 40 days. The models were initialised using the ECMWF 9 km analysis. Daily observed sea-surface temperatures and sea ice concentrations were used as boundary conditions. Top of atmosphere outgoing longwave radiation (OLR) is available for each model simulation, regridded to a consistent 0.1 degree grid. We discard the first ten-days of each simulation as `spin-up' as suggested in \citet{stevens2019}.

The DYAMOND protocol was deliberately simple and unprescriptive, to encourage as many groups to participate as possible. The model simulations therefore differ in their choice of vertical resolution, model top, cumulus parametrisation scheme, and boundary layer scheme, among many other modelling choices \citep{stevens2019}. The linear resolution (square root of the maximum grid box size) ranges from 2.5 km to 7.8 km for the core ensemble. In addition to the core ensemble, we analyse six simulations from the `incidental ensemble'; these simulations differ from the core simulations in their horizontal resolution and choice of cumulus parametrisation, allowing us to test the sensitivity of fractal dimension to these modelling choices.

\subsection{Observed cloud fields}

We compare the global storm-resolving simulations to satellite derived cloud fields. The Japan Meteorological Agency Himawari 8 satellite was the most sophisticated geostationary satellite in orbit in 2016, and was used in \citet{stevens2019} to validate the DYAMOND simulations: we also use Himawari 8 data for consistency with that paper. Himawari 8 has been operational since July 2015 \citep{bessho2016}. The satellite is positioned at $140.7^\mathrm{o}$E, with local solar noon at 02:41 UTC. The derived cloud-top temperature and total cloud optical thickness products used in this analysis are provided on a 5 km grid, with 10 minute resolution, in daytime regions only.

\subsection{Region}

We restrict our attention to an area below the footprint of the Himawari satellite. Throughout the analysis we will approximate pixels as squares. This limits the latitudinal extent of the domain we can consider. We use data between $25^{\mathrm{o}}\mathrm{N}$ and $25^{\mathrm{o}}\mathrm{S}$, such that the maximum fractional distortion in the longitudinal arclength is $\mathrm{cos}(25^\mathrm{o}) = 0.91$. The longitudinal range is $80-180^{\mathrm{o}}\mathrm{E}$. 

\section{Methods} \label{sec:methods}

\subsection{Defining cloud objects}

We follow the methodology presented in \citet{lovejoy1982}. We first compute a binary cloud field from the observed cloud-top temperature data using a threshold of 215 K: any pixel with cloud-top temperature below this threshold is defined as a cloudy pixel. This selects cold, deep clouds. Setting the threshold colder restricts the number of clouds analysed, while choosing a warmer threshold makes it harder to compare satellite to model data. A warmer threshold also reduces the number of clouds, because cloud objects that intersect the edge of the domain are discarded. We found our analysis was robust to changes to this threshold --- indeed, our observed estimate of the fractal dimension is not significantly different to that reported in \citet{lovejoy1982}, despite our chosen thresholds differing by 50 K.

To compute equivalent binary cloud fields for each simulation, we must map between cloud-top temperature and OLR fields. This is carried out in a  two-stage process. First, we map between the OLR and the effective radiating temperature of the cloud, $T_{rad}$. The relationship is not simply that of a black body flux, because of absorption of radiation by the atmosphere above cloud top. We therefore use the empirically measured relationship of \citet{VaillantDeGuelis2017},
\begin{equation}
   F_{TOA} = \alpha T_{rad} + \beta \label{eq:OLR_to_Trad}
\end{equation}
where the OLR, $F_{TOA}$, is in $\mathrm{W m}^{-2}$, $\alpha = 2.0 \, \mathrm{W m^{-2} K^{-1}}$ and $\beta = -310 \, \mathrm{W m^{-2}}$. Note that this linear relationship between OLR and cloud radiating temperature was originally identified by \citet{ramanathan1977}.

The effective radiating temperature of the cloud will typically reflect the temperature some depth below the cloud-top. To convert between cloud radiating and cloud top temperature, we compare the satellite cloud-top temperature field to the cloud radiating temperature derived from one of the DYAMOND simulations following equation~\ref{eq:OLR_to_Trad}. We use the ICON 2.5 km simulated field for this mapping, regridded to the same 5 km grid as the satellite data. We use model data from 15 minutes after initialisation to ensure a strong equivalence between the simulated and observed cloud fields. Despite the short lead time, there are likely to be cloudy pixels in the model simulation where no cloud was observed in satellite data, and vice versa. To ensure we only compare locations where there is a high, thick cloud in both model and satellite data, we only compare pixels for which the Himawari cloud optical thickness is at least 2.5 and the ICON column integrated ice and water content exceeds $0.005 \, \mathrm{kg \, m}^{-2}$ and $0.0 \, \mathrm{kg \, m}^{-2}$ respectively. A total least squares fit on the retained pixels produced
\begin{equation}
    T_{rad} = 0.86 T_{CT} + 36.1 \label{eq:TCT_to_Trad}
\end{equation}
where the cloud-top temperature, $T_{CT}$, and radiating temperature, $T_{rad}$, are in K. Combining equations ~\ref{eq:OLR_to_Trad} and ~\ref{eq:TCT_to_Trad}, the cloud-top temperature threshold of 215 K corresponds to a threshold in OLR of 132 $\mathrm{Wm}^{-2}$. We use the same OLR threshold for all model simulations.

\subsection{Area-Perimeter fractal dimension}

The fractal dimension of an object can be measured in several different ways depending on the nature of the object in question, including the similarity dimension \citep{strogatz} or box-counting dimension \citep{Walsh1993}. For two-dimensional objects, the area-perimeter relation is a natural choice \citep{lovejoy1982}. This relation predicts a fractal dimension of the perimeter through the equation $P \propto A^{D/2}$. The fractal dimension, $D$, thus derived is a measure of the complexity, or `crinkliness', of the perimeter, since a smooth perimeter of given length encloses more area than a complex one. 

We compute an instantaneous fractal dimension for each observed cloud top temperature and simulated OLR field. After applying the relevant threshold to generate a binary cloud field, the images are cleaned by removing holes in each cloud object. We discard any clouds with size smaller than 24 pixels as it is not possible to accurately estimate the cloud perimeter for objects of this size and smaller. Note that this pixel threshold means that the lower resolution model fields have a physically larger minimum size of cloud than the satellite fields. We finally remove any clouds which overlap the edge of the field, as the perimeters of these clouds would be anomalously smooth.

\section{Results} \label{sec:results}

\subsection{Fractal dimension of cloud objects}

\begin{figure}
\centering
\includegraphics[width=\textwidth,trim={0.0cm 0 0.0cm 0},clip]{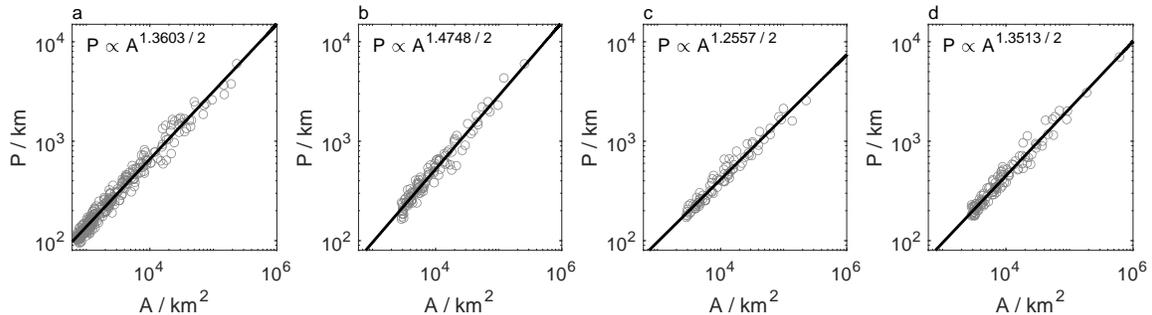}
\caption{Cloud object perimeter as a function of area for a) Himawari satellite data, and for b) ICON 2.5 km, c) IFS 4.8 km, and d) NICAM 3.5 km simulations. The data correspond to fields at 0200 UTC on 11 August 2016.}\label{fig:scatterD}
\end{figure}

The area and perimeter of each cloud object in each binary field is measured. Figure~\ref{fig:scatterD}(a) shows the measurements from 0200 UTC on 11 August 2016 of observed cloud objects identified from the satellite cloud top temperature field. As demonstrated in \citet{lovejoy1982}, the observed cloud objects are fractals, exhibiting the expected scaling behaviour. Linear regression provides an estimate of the instantaneous fractal dimension, D, of 1.36, consistent with the value of 1.35 reported in \citet{lovejoy1982}. Figures~\ref{fig:scatterD}(b-c) show the equivalent area-perimeter measurements for simulated cloud fields from three of the models. We find that the cloud objects in the storm-resolving simulations also exhibit fractal behaviour, showing a clean scaling relation similar to that observed in satellite data. The instantaneous fractal dimensions computed for each simulated field are close to the estimate from the satellite-derived image. The linear relationship shown for observed and simulated fields supports monofractal behaviour, as opposed to a multifractal scaling law. 

\begin{figure}
\centering
\includegraphics[width=\textwidth]{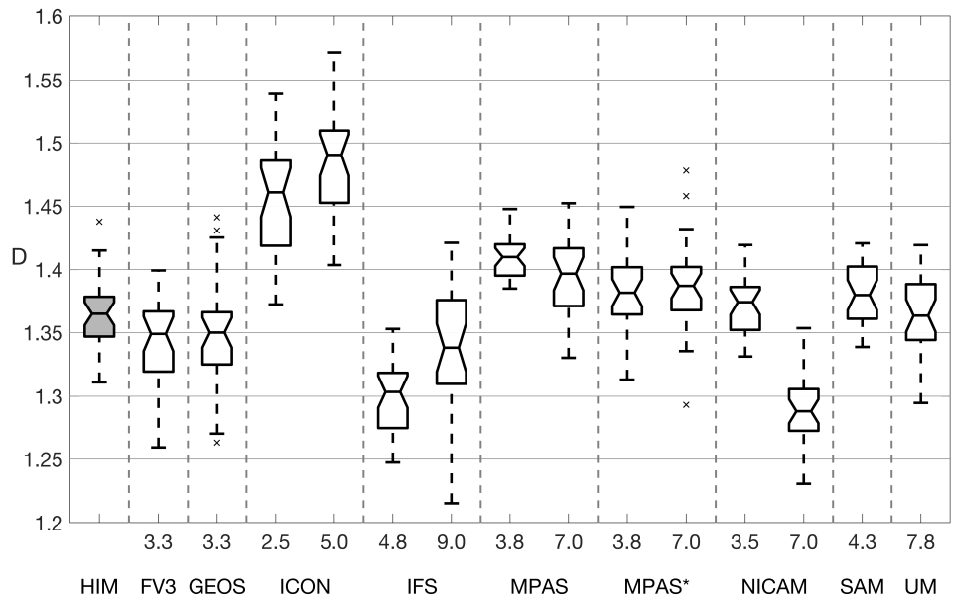}
\caption{Box and whisker plot showing the distribution of fractal dimension, D, computed for Himawari satellite data (HIM: grey box plot) and each model simulation (white box plots). The x-axis labels indicate the model and the resolution of each simulation in km. The MPAS* simulations use the full convection parametrisation, identical to that used in coarser MPAS simulations. Each box top, waist, and bottom correspond to the 75th, 50th and 25th percentiles respectively. The whiskers indicate the range of the distribution. Crosses indicate outliers, defined as values that are more than 1.5 times the interquartile range away from the bottom or top of the box. The data correspond to 0200, 0300, and 0400 UTC for the last 30 days of the simulation (11 August--9 September 2016).}\label{fig:boxplot}
\end{figure}

The fractal dimension is computed separately for each day between 11 August and 9 September 2016 as the average of the dimension computed for 0200, 0300 and 0400 UTC. These times are close to local noon, for which satellite estimates of cloud top temperature are available. This analysis is performed for the satellite data and for each model simulation. The distribution of the computed fractal dimension $D$ over this 30-day period is indicated in Figure~\ref{fig:boxplot}. All the global storm-resolving models simulate cloud fields that exhibit fractal behaviour. The fractal dimension measured for the simulated cloud fields varies from model to model. For example, the ICON 2.5 and 5 km simulations have a notably higher fractal dimension than the other models (i.e. they have more crinkly clouds), while the IFS 4.8 km and NICAM 7 km simulations have a notably lower fractal dimension (i.e. they have smoother clouds). However, in general, the fractal dimension of the simulated cloud fields are all remarkably similar to that measured using satellite data, with the median dimension for all models falling within 10\% of the observed value. 

\begin{figure}
\centering
\includegraphics[width=0.7\textwidth,trim={0.0cm 0 0.0cm 0},clip]{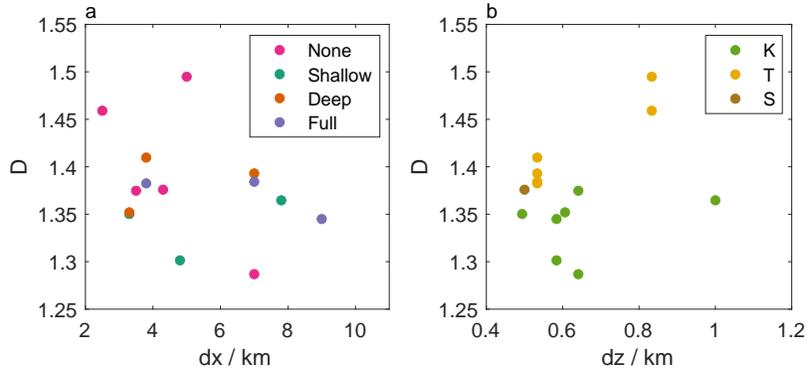}
\caption{The fractal dimension, D, as a function of (a) the horizontal resolution and (b) the average vertical resolution, $\mathrm{d}z = H_{top}/n_{lev}$. In (a), the marker colours indicate the cumulus parametrisation scheme used: `none', `shallow', `deep', or `full', where the `deep' scheme has been re-tuned or adjusted to account for the scales of motion being parametrised, whereas `full' is the un-tuned version also used at coarser resolutions. In (b) the marker colours indicate the boundary layer scheme used: `T', a TKE-type scheme, involving an additional prognostic equation; `K', a diagnostic eddy diffusivity scheme; `S', a three-dimensional Smagorinsky-type scheme. The data correspond to 0200, 0300, and 0400 UTC for the last 30 days of the simulation (11 August--9 September 2016). }\label{fig:predictor}
\end{figure}

Within the DYAMOND ensemble, the model set up differs substantially between simulations \citep{stevens2019}. We consider whether any of the key choices made by the modelling groups contribute to the observed differences in simulated fractal dimension. Figure~\ref{fig:predictor} shows the relationship between fractal dimension and the model resolution in both the horizontal and vertical. A small negative correlation (-0.32) is measured for horizontal resolution, while a small positive correlation (0.35) is measured for vertical resolution, though neither correlation is significant at the 95\% level. The colour of the data points in Figure~\ref{fig:predictor} also indicates the choice of convection scheme and boundary layer scheme. Interestingly, there is no systematic relationship between the choice of convection scheme and the simulated fractal dimension. However, sensitivity is observed in the choice of boundary layer scheme. Models which use a turbulent kinetic energy (TKE) scheme, in which closure involves solving a prognostic equation for TKE, systematically have a higher fractal dimension than models which use a diagnostic eddy-diffusivity scheme. One simulation analysed uses a 3-dimensional Smagorinsky-type scheme, as is commonly used in Large Eddy Simulations: the fractal dimension of cloud fields in that simulation falls between the dimension estimated for the models using a TKE or eddy-diffusivity scheme. We also considered the role of the height of the model top, the height of the sponge layer, and the number of vertical levels: none of these showed a significant impact on the simulated fractal dimension. 

\begin{figure}
\centering
\includegraphics[width=\textwidth]{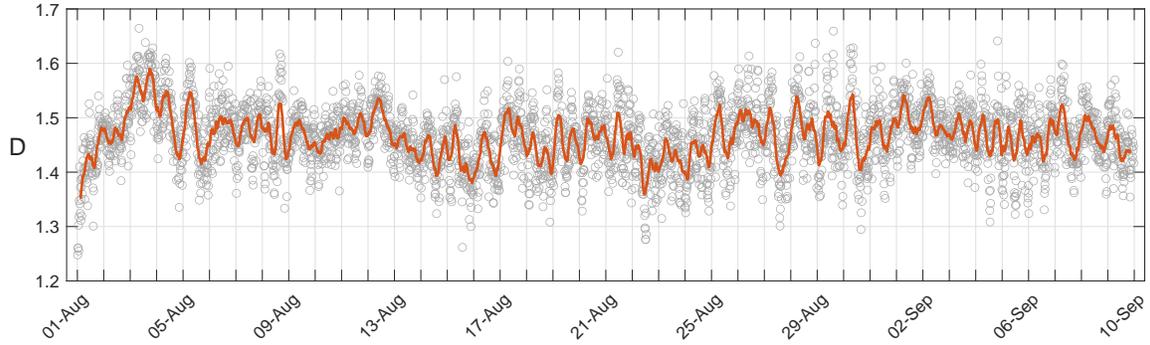}
\caption{Time series of the fractal dimension of cloud objects in the ICON 2.5 km simulation. The scattered points are instantaneous measurements while the red line shows a 6-hr rolling average.}\label{fig:time}
\end{figure}

While the satellite derived binary cloud fields are only available during daylight hours, the simulated cloud fields are also available during the night. Figure~\ref{fig:time} shows the fractal dimension computed for the ICON 2.5 km simulation as a function of time, including both instantaneous values and a 6-hour rolling average. Here we consider the entire 40-day simulation. An initial spin-up period of 4-5 days is evident, during which the fractal dimension increases before converging on a stable value. This is shorter than the spin up period of ten-days highlighted by \citet{stevens2019} for these simulations, suggesting ten-days is overcautious. Substantial variability is observed in the fractal dimension. This includes both a diurnal cycle and lower frequency variations. 

\subsection{Distribution of cloud areas}

\begin{figure}
\centering
\includegraphics[width=0.8\textwidth]{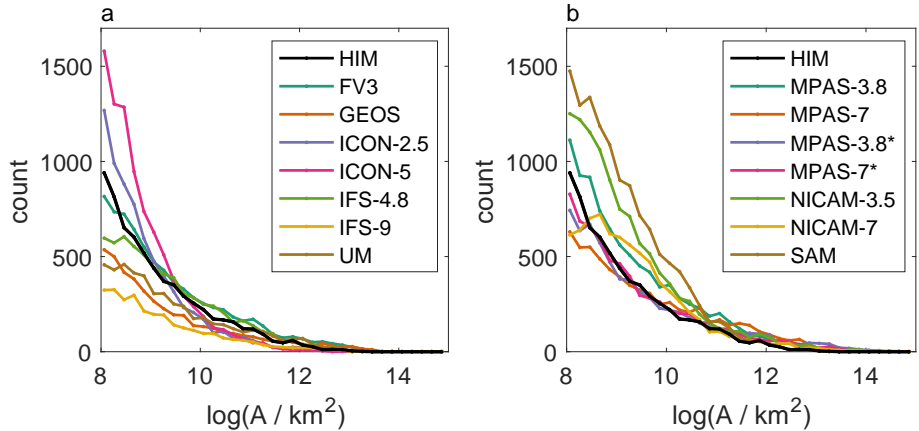}
\caption{Number of cloud objects as a function of cloud object area, A, compared to Himawari satellite data (black). The simulations (colours) are split into two subsets (a) and (b) for clarity. The data correspond to 0200, 0300, and 0400 UTC for the last 30 days of the simulation (11 August--9 September 2016). }\label{fig:pdf}
\end{figure}

We finally measure the distribution of cloud object areas. This complementary diagnostic also provides an independent assessment of whether the OLR and cloud top temperature thresholds are consistent with each other. Figure~\ref{fig:pdf} shows the number of cloud objects as a function of cloud object area for the satellite and model data sets. On average across the models, the simulated distributions match the observed distribution, suggesting the thresholds in OLR and cloud top temperature are indeed consistent. However there are substantial differences between simulations produced using different models. When the same model is used but the resolution is varied, the resolution tends to have a large impact on the simulated distribution, except for the MPAS* simulations. This pair of simulations are from the `incidental ensemble' and include the same version of the MPAS convection scheme that is used at coarser resolutions. This indicates the full MPAS convection scheme is scale aware. 

We considered a range of metrics to summarise the fidelity of the distribution of simulated cloud objects captured in figure~\ref{fig:pdf}, including the average total area of cloud objects, and measures of the difference between modelled and observed cloud distributions. We found no correlation between any of these metrics and the fractal dimension (not shown): these two diagnostics are independent of each other.

\section{Discussion and Conclusions} \label{sec:conc}

We describe a novel methodology for examining the fidelity of cloud fields in high resolution model simulations. We define a binary cloud field using a threshold in OLR or cloud top temperature, before examining the fractal nature of the resultant cloud objects. We demonstrate that global storm-resolving simulations produced for the DYAMOND project \citep{stevens2019} are able to reproduce the observed fractal nature of clouds. In other words, the area and perimeter of cloud objects are related by the scaling law $P \propto A^{D/2}$ where D is the fractal dimension. The fractal nature of clouds is an emergent property of models. It gives us confidence that these global storm-resolving models are representing the key aspects of the physics of clouds, convection, and convective organisation and aggregation. 

In addition to reproducing the fractal nature of clouds, the fractal dimension of clouds in these simulations is close (within 10\%) to that of cloud fields observed by the Himawari 8 geostationary satellite. To explain differences in the fractal dimension between models, we considered the modelling choices made when producing each of the DYAMOND simulations. The different simulations vary substantially in their model setups, including the model horizontal and vertical resolutions, the type of boundary layer scheme used, and whether any part of the model's convection parametrisation scheme is activated. Of these choices, the only significant predictor of fractal dimension was the choice of boundary layer scheme. Models which included a prognostic equation for TKE had a higher fractal dimension than those which use an eddy diffusivity scheme. This highlights the importance of the boundary layer scheme for convective organisation \emph{as opposed to the convection scheme}. The comparative role of these two parametrisation schemes in organising convection is poorly known, though some single-model studies have highlighted the importance of the choice of boundary layer scheme for phenomena such as the Madden Julian Oscillation (MJO) \citep{holloway2013}, consistent with the Convective-Dynamic-Moisture trio-interaction theory of the MJO \citep{zhang2020}. More generally, it is known that the different approximations used in boundary layer parametrisations can significantly impact the climate in models, including the model's climate sensitivity \citep{Garratt1993,Davy2014,Davy2016}.

We noted substantial temporal variability in the fractal dimension of clouds as simulated by a single model. Future work will seek to understand this variability, including aspects related to the diurnal cycle and lower frequency modes, as an indicator of changes in convective organisation.

One of the motivations behind this study was to quantify the `Palmer-Turing test' \citep{Palmer2016}. This test suggests that we can assess the fidelity of climate models by visually comparing instantaneous maps of model output to satellite observations. If we cannot tell which image is simulated and which is observed, then the model has passed the Palmer-Turing test. We suggested that the fractal dimension of cloud objects would be a useful tool to quantify and explain the visual similarities and differences noted when performing the Palmer-Turing test, which are otherwise difficult to enunciate. It is instructive to return at this point to the images of the simulated condensate fields from the DYAMOND simulations \cite[Figure 2]{stevens2019}. The model whose instantaneous fractal dimension is closest to Himawari at 0400 UTC on 4 August 2016 is FV3 (D = 1.38 compared to the observed D=1.37), followed by NICAM and the UK Met Office. We note that these three models are visually different to each other. The fractal dimension alone does not fully characterise the image. In particular, low cloud and the distribution of cloud object sizes also influence our impression. Nevertheless, the fractal dimension is a useful additional tool for quantifying the ability of models to simulate deep convection and convective organisation.

%
%
%
%
%
%
%
%

\section*{Acknowledgments}

The authors thank Simon Proud for assistance with the Himawari 8 data, and Daniel Klocke and Florian Ziemen for assistance with the DYAMOND data. The processed cloud data used to compute the fractal dimension in this study can be downloaded from DOI:10.5281/zenodo.5196327 under a GNU general public licence v2.0.
H.M.C. was funded by Natural Environment Research Council grant number NE/P018238/1. 
DYAMOND data management was provided by the German Climate Computing Center (DKRZ) and supported through the projects ESiWACE and ESiWACE2. The projects ESiWACE and ESiWACE2 have received funding from the European Union’s Horizon 2020 research and innovation programme under grant agreements No 675191 and 823988. This work used resources of the Deutsches Klimarechenzentrum (DKRZ) granted by its Scientific Steering Committee (WLA) under project IDs bk1040 and bb1153.


%
%

\bibliography{Christensen_Driver_Fractal}

\bibliographystyle{agsm}



%


\end{document}